\documentclass[useAMS,usenatbib,usegraphicx]{mn2e}
\usepackage{mathptmx}
\usepackage{color}
\usepackage{txfonts}
\usepackage{graphicx}
\usepackage{longtable}
\usepackage{lscape}
\usepackage{pdflscape}

\title{The chemical case for no winds in dwarf irregular galaxies}

\author[Gavil\'{a}n et al.]
{
Marta Gavil\'{a}n$^1$, Yago Ascasibar$^1$, Mercedes Moll\'{a}$^2$, and \'{A}ngeles I. D\'{i}az$^1$\\
$^1$ Departamento de F\'{\i}sica Te\'{o}rica, Universidad Aut\'onoma de Madrid, 28049 Madrid, Spain\\
$^2$ Departamento de Investigaci\'{o}n B\'{a}sica, C.I.E.M.A.T., Avda. Complutense 22, 28040 Madrid, Spain\\
}

\date{{\bf Draft version 3.3}, \today}

\newcommand{\msun}{\ensuremath{{\rm M_\odot}}}

\newcommand{\aap}{A\&A}

\newcommand{\aj}{AJ}
\newcommand{\apj}{ApJ}
\newcommand{\apjl}{ApJ}
\newcommand{\apjs}{ApJS}

\newcommand{\araa}{ARA\&A}
\newcommand{\mnras}{MNRAS}

\begin{document}
\maketitle

\begin{abstract}

We argue that isolated gas-rich dwarf galaxies -- in particular, dwarf irregular (dIrr) galaxies -- do not necessarily undergo significant gas loss.
Our aim is to investigate whether the observed properties of isolated, gas-rich dwarf galaxies, not affected by external environmental processes, can be reproduced by self-consistent chemo-photometric infall models with continuous star formation histories and no mass or metals loss.
The model is characterized by the total mass of primordial gas available to the object, its characteristic collapse timescale, and a constant star formation efficiency.
A grid of 144 such models has been computed by varying these parameters, and their predictions (elemental abundances, stellar and gas masses, photometric colours) have been compared with a set of observations of dIrr galaxies obtained from the literature.
It is found that the models with moderate to low efficiency are able to reproduce most of the observational data, including the relative abundances of nitrogen and oxygen.
\end{abstract}

\begin{keywords}
galaxies: dwarfs -- galaxies:formation -- galaxies: evolution -- galaxies: abundances
\end{keywords}

\maketitle

\section{Introduction}

According to the hierarchical model of structure formation, the first objects to form in the universe are the smallest galaxies, and these systems act as the building blocks of larger structures. Dwarf galaxies, defined by their low luminosities (absolute magnitude larger than -18) and small sizes (diameter $< 10$ kpc), play a very important role in such a \textit{bottom-up} scenario, and their study is absolutely essential in order to understand galaxy formation and evolution at all scales and epochs.

Present-day dwarf galaxies can be broadly classified into gas-rich and gas-poor objects. Gas-poor dwarf galaxies, such as dwarf ellipticals (dE) and spheroidals (dSph), are usually found in crowded environments, while gas-rich galaxies, like dwarf irregulars (dIrr) and blue compact dwarfs (BCD) seem to prefer low-density environments, mostly in the field but sometimes even in the outer parts of galaxy groups and clusters.  It has been argued that the formation of all dwarf galaxies should be similar, and the differences between gas-rich and gas-poor dwarf galaxies arise only during their evolution, but the observational evidence is far from being conclusive \citep[see e.g.][and references therein]{ski95}.
If true, a mechanism like ram-pressure stripping \citep{ken89, cay94,rij10} for the gas-poor galaxies, or gas accretion during the life of the galaxy \citep{lac85,moo94,mou02,kam08} for the gas-rich ones, must be invoked in order to explain why 
low-mass objects have less gas in dense and hot environments than in the field. Besides that, recent observations \citep[e.g.][]{dun10} indicate that dEs/dSph and dIrr constitute two structurally distinct populations.

In this work we focus our attention on the evolution of isolated gas-rich dwarf galaxies, not affected by external environmental processes.
One of the agents that may regulate the star formation history of these objects is the presence of strong galactic winds.
If they have a low mass and a high star formation rate (SFR), the energy injected by supernova explosions may be able to eject some fraction of their gas into the intergalactic medium \citep{dek86}.
The existence of such an outflow would be critical for the evolution of dwarf galaxies, and it has produced a lot of work in the last decades.
In particular, it is still debated whether (or to what extent) dwarf galaxies are able to retain their gas in spite of their shallow potential wells, as well as the possible links between galactic winds and the physical properties of the different dwarf galaxy types.

\citet{ski97b} made a clear review of the arguments for and against galactic winds dominating the evolution of dIrr galaxies. Favouring the idea of moderate winds, gas flows with high velocities have been observed in several objects, like NGC\,1569.
However, a search for ionized hydrogen in the outer parts of 51 dwarf galaxies by \citet{hun93} yielded only 2 positive results. More recent works by \citet{van07,van09a,van09b,van10} have detected outflows with velocities in the range $20-60$ km~s$^{-1}$ in several dwarf galaxies. These values  suggest expanding super-bubbles, but they are not sufficient to allow the gas to escape from the gravitational potential well.

From a chemical evolution point of view, the low oxygen abundances found in many dIrr do not seem to be well reproduced by the simple closed-box model (SCBM).
A possible solution to this problem would be to invoke the presence of enriched galactic winds, ejecting part of the metals produced during the early star formation episodes.
However, if one assumes that a large fraction of the oxygen produced in the first stages blows out of the galaxy, the observed values of the C/O relative abundances cannot be explained \citep{car95}.

The situation with the N/O ratio is even more critical; according to \citet{lar01}, selective winds do not give a good fit to N/O, while closed models and models with non-selective winds -- with or without inflow -- are all found to be viable. On the other hand, other authors \citep{rec01,chi03-2,lan03,rom06,yin11} argue that models with strong and/or enriched winds are able to reproduce all the available observational constraints.

It is important to remark that most of the studies that suggest the existence of galactic winds \citep[e.g.][]{meu92,pap94,mar95} are based on observations of bursting galaxies.
While this is often the case in BCDs (and, in fact, several authors include a minimum equivalent width of the H$\alpha$ emission line among the selection criteria), dIrr, which constitute \emph{the vast majority} of dwarf galaxies, display much more moderate star formation rates (see the corresponding column in Table~\ref{tabla_dIrr}).
Roughly speaking, we identify the bursting and quiet populations with BCD and dIrr, respectively.
Of course, real galaxies will probably form a continuum between these two extremes, but it is our aim to show that current observational data of dIrr galaxies are consistent with a scenario where most of these objects form relatively quietly and never develop any significant outflow.
Such a scenario would also be consistent with the number statistics of starbursts in the local Universe: according to \citet{Lee09}, only six percent of the local dwarf galaxies are currently undergoing a starburst phase, and the fraction of stars formed in that mode is of the order of 25 per cent.

According to many theoretical studies \citep{mac99,der99,rec01,fra04} galactic winds do not expel a significant fraction of gas.
The outflows occur in a direction normal to the disk of the galaxy, along which the pressure gradient is steeper, and there is little horizontal transport.
The ejection efficiencies of the metals synthsized by Type~II supernovae  (e.g. oxygen) can be quite high, since they can be easily channeled out of the galaxy, but the ejection efficiency of pre-existing gas would generally be very low. Elements produced in low- and intermediate-mass stars, such as nitrogen and carbon, could still be channelled along the tunnel carved by Type~II supernovae, although it is unclear whether the winds of planetary nebulae would be strong enough to overcome the gravitational potencial dominated by the dark matter halo. In the case of long-lasting star formation, the required energy could be provided by SN explosions.
Theoretical efforts made in order to assess the relevance of galactic winds can be divided into two main groups: one approach, mostly based on hydrodynamical simulations \citep[e.g.][]{mac99,sil01}, focuses on the energy requirements for developing efficient galactic outflows, while the other is more concerned with chemical evolution and is often based on the analysis of purely chemical \citep{mat85,lar01,fer00,rom06} or spectrophotometric models \citep{fri94}, or a combination of both \citep{boi03,mar08}.

Unfortunately, it is very difficult to join these two approaches in a self-consistent way, although some efforts are being made in this direction. In the last years, numerical techniques have drastically improved, and one may follow the full dynamics of the galaxy by solving the Euler equation, including also a chemical evolution scheme  \citep[e.g.][]{rec01, mar08b,val08,rev09,saw10}.
In general, these numerical simulations focus on the hierarchical assembly of the galaxy and the cooling of the intergalactic gas.
Important physical processes, such as the formation of molecular clouds, the presence of magnetic fields, cosmic rays, small-scale turbulence, as well as the effects of the ionizing and dissociating radiation from the stars, are usually not taken into account.
Therefore these studies  favour very strong winds, with star formation histories characterized by repeated cycles of bursts and periods of moderate star formation activity \cite[e.g.][]{sti07}. Available observations in the dwarf galaxies of the Local Group and in some other nearby dwarf galaxies -- see for instance the review of \cite{tol09}, or the recent results of the LCID project -- seem to support this `breathing' star formation scenario.

Here we explore the possibility that some unidentified physical process (e.g. photodissociation) lowers the efficiency of star formation without driving the gas out of the galaxy, in agreement with the most recent observations \citep[see][]{bom07,van07,van09a,van09b,van10} as well as several theoretical studies \citep[e.g.][]{leg01,tas08}.
More detailed modelling (ideally, an ab-initio cosmological simulation) would be required in order to pinpoint the specific process(es) that are actually responsible for regulating the star formation efficiency.

More precisely, our aim is to investigate whether the observed properties of isolated, gas-rich dwarf galaxies, not affected by external environmental processes, can be reproduced by self-consistent chemo-photometric infall models with continuous star formation histories and no mass loss.
In particular, we study in detail the composition of the gaseous phase, the abundance ratios of different chemical elements, and their relation with other properties, such as gas fractions and photometric colours.
In other words, we address the question of whether/when the presence of strong galactic outflows is \emph{necessary} in order to simultaneously reproduce an updated set of photometric and chemical data \citep[among which we would like to highlight the N/O ratio; see e.g.][]{lar01,mol06}.

Our models are described in Section~\ref{secModels}, and Section~\ref{secResults} is devoted to the comparison of their predictions with a compilation of observational data selected from the literature.
A succinct summary of our main conclusions is provided in Section~\ref{secConclusions}.

\section{Theoretical models}
\label{secModels}

The basic tenet of our model is that dwarf irregular galaxies are forming continuously throughout their lifetime, in the sense that infall of primordial gas into the galaxy and star formation at some level are taking place at all times. For disk galaxies, this idea was proposed a long time ago: since the contribution of \citet{lar72} and \citet{lyn75}, there have been many authors that worked on this scenario to explain the G-dwarf problem and the abundance gradients seen in our Galaxy. This is the case of the chemical evolution model designed by \citet{fer92}, later updated by \citet{mol05}, that we have adapted for dwarf galaxies in the present work.

The main features of the model are:
1) The galaxy is modelled as one single region.
2) There is a continuous infall of primordial gas from the surrounding medium into the galaxy.
3) There are no winds or outflows that remove gas from the galaxy.
4) The star formation process is regulated by the amount of gas available at each moment.

In the present work, the system is modelled as an isolated, single region, without any substructure.
While this is obviously insufficient to describe any property that is related to spatial distribution, it certainly suffices to investigate the integrated properties \citep[e.g.][]{yin11}.
The model starts from a primordial region of the universe (protogalaxy) that slowly collapses onto the central part (galaxy). The infall rate is set by the characteristic collapse time $\tau$, which is a free parameter of the model\footnote{This prescription admits two physically different (yet mathematically equivalent) interpretations: a) the gas falls at very high redshift and forms a hot halo. $\tau$  represents the time it takes the gas to cool down, collapse gravitationally, and form the atomic phase of the ISM; b) $\tau$ is the time it takes the intergalactic gas to fall into the galaxy. The timescale to condense into the atomic phase is much shorter, and there is virtually no hot halo.}.
Within the galaxy, molecular clouds are formed with an efficiency $\epsilon_{\mu}$, and a proportion of them transforms into stars according to an efficiency parameter $\epsilon_{h}$. 
Five different phases of matter are considered: diffuse gas, molecular gas, massive stars, low and intermediate mass stars and remnants. The distinction between massive and intermediate mass stars is due to nucleosynthesis prescriptions.

The model equations, defined in \cite{fer94}, are
\begin{eqnarray}
 \frac{ds_{1}}{dt}&=&H_{1} c^{2} + a_{1} c s_{2}  -D_{1}\\
 \frac{ds_{2}}{dt}&=&H_{2}c^{2} + a_{2} c s_{2} - D_{2}\\
 \frac{dr}{dt}&=&D_{1}+D_{2}-W\\
 \frac{dg}{dt}&=&-\mu g^{n}+a'cs_{2}+H'c^{2}+ \frac{ g_{ext} }{ \tau } + W\\
 \frac{dg_{\rm ext}}{dt} &=& - \frac{ g_{ext} }{ \tau }\\
 \frac{dc}{dt}&=&\mu g^{n}-\left(a_{1}+a_{2}+a'\right)cs_{2}-\left(H_{1}+H_{2}+H'\right)c^{2}\\
 \frac{dX_{i}}{dt}&=& \frac{ W_{i} - X_{i}W  + \frac{ g_{ext} }{ \tau } ( X_{i, ext} -  X_{i} ) }{ g + c }
\end{eqnarray}
where
\begin{itemize}
\item $s_1$ = Stellar mass (stars more massive than 8~\msun)
\item $s_2$ = Stellar mass (stars less massive than 8~\msun)
\item $r$ = Mass in stellar remnants
\item $g$ = Gas mass in the galaxy
\item $g_{ext}$ = Gas mass in the external reservoir
\item $c$ = Mass in molecular clouds
\item $X_i = \frac{M_i}{g+c}$ = Relative abundance of element $i$ ($^{1}$H, D, $^{3}$He, $^{4}$He, $^{12}$C, $^{16}$O, $^{14}$N, $^{13}$C, $^{20}$Ne, $^{24}$Mg, $^{28}$Si, $^{32}$S, $^{40}$Ca, $^{56}$Fe,  and the combined abundance of neutron-rich isotopes of C, N, and O)
\end{itemize}
and the equation coefficients correspond to different physical processes:
\begin{enumerate}
 \item Gas accretion from the protogalaxy: $\frac{ g_{ext} }{ \tau }$
 \item Cloud formation from diffuse gas: $\mu g^{n}$, with $n= 1.5$
 \item Star formation due to cloud collisions: $H_{1,2}c^{2}$
 \item Diffuse gas restitution due to cloud collisions: $H'c^{2}$
 \item Induced star formation due to the interaction between clouds and massive stars: $a_{1,2}cs_{2}$
 \item Diffuse gas restitution due to the induced star formation: $a'cs_{2}$
 \item Stellar death rate: $D_{1,2}$
 \item Fraction of mass injected as element $i$: $W_i$
 \item Total restitution rate: $W=\sum W_i$
\end{enumerate}

Although the number of parameters seems to be large, actually not all of them are free. For example, ${\rm H_{1,2}}$ and ${\rm a_{1,2}}$ are estimated by the equation
\begin{equation}
 \Psi(t)=(H_{1}+H_{2})c^{2}+(a_{1}+a_{2})cs_{2}
\end{equation}
while the efficiencies result from the combination of the probability of cloud formation, $\epsilon_{\mu}$, and cloud collision, $\epsilon_{H}$ \citep[see][for further details]{mol05}.
Thus, the number of free parameters is reduced to three: $\tau$, $\epsilon_{\mu}$ and $\epsilon_{H}$.


It is important to stress that the SFR is not an input of the model, but a result of the efficiencies selected and the amount of gas available. The higher the efficiencies, the higher the SFR will be, until the gas reservoir in the protogalaxy is completely exhausted. The ensuing star formation histories are smooth and continuous, but not constant. The SFR will have different values through time, although sudden variations (i.e. bursts) never occur in our models.


\begin{table}
\caption{Adopted values for the total mass of the protogalaxy and the galactic radius, as defined in Moll\'{a} \& D\'{i}az (2005) }
\begin{center}
\begin{tabular}{|c|c|c|}
\hline
Model &   Total mass   & Galaxy radius \\
      & [$10^6$ \msun] &     [pc]      \\
\hline
M1 &   10 &  96 \\
M2 &   50 & 151 \\
M3 &  100 & 184 \\
M4 &  500 & 290 \\
M5 & 1000 & 353 \\
M6 & 5000 & 557 \\
\hline
\end{tabular}
\label{tabla_Protogalaxy}
\end{center}
\end{table}

\begin{table}
\caption{Values of the star formation efficiency parameters.}
\begin{center}
\begin{tabular}{|c|c|c|}
\hline
Efficiency& ${\rm \epsilon_{\mu}}$ &  ${\rm \epsilon_{h}}$\\ 
\hline
Very low &6.74$\times10^{-3}$&3.73$\times10^{-6}$ \\ 
Low&0.0174& 4.01$\times10^{-5}$\\ 
Medium-low&0.0408& 3.36$\times10^{-4}$\\\ 
Medium&0.165&0.0111 \\ 
High&0.638&0.325 \\ 
Very high&0.951& 0.883 \\
\hline 
\end{tabular} 
\label{tabla_Efficiencies}
\end{center}
\end{table}

We have constructed a grid of models with different values of the total mass, collapse time, and star formation efficiencies. The dynamical masses and the galaxy radii are selected extrapolating \cite{mol05} models toward lower masses. We have selected total protogalaxy masses between $10^7$ and $5\times10^9$~\msun (actual values are quoted in Table~\ref{tabla_Protogalaxy}). For each value, different models have been run with collapse times equal to 8, 20, 40 and 60~Gyr, eliminating the link between this parameter and the total mass of the galaxy existing in \cite{mol05} models, since for this mass range that relation produces unrealistically long collapse times scales. This will have a small effect on the star formation rate through the slightly different densities of the galaxies, but the dominant factor that determines the SFR is the chosen efficiency.
In order to reduce the number of free parameters, the efficiencies in forming clouds and stars have been grouped together, and we consider six different sets, from ``very low'' to ``very high'' values (see Table~\ref{tabla_Efficiencies}) in the range [0,1]. "Very low" and "very high" efficiencies correspond to \cite{mol05} N=10 and N=1 models, while the other values are interpolated between them. In total, our grid is composed of 144 models, corresponding to 6 values of the total mass, 4 different collapse times, and 6 star formation efficiencies.

We build the spectral energy distribution (SED) of each model galaxy by convolving the SEDs of individual single stellar populations (SSP) with the predicted star formation and chemical enrichment history of the object.
Luminosities in different photometric bands are then computed by applying the appropriate filters to the galaxy SED \citep{gir04}.
A single stellar population is composed of stars of different masses, according to a certain IMF, that are formed in a short period of time, so they can be considered to have the same age and metallicity.

In our case, their spectral energy distributions have been obtained from the models by \citet{mol09}, based on the updated isochrones from \citet{bre98} for 6 different metallicities: $\rm Z= 0.0001$, 0.0004, 0.004, 0.008, 0.02 and 0.05. The age coverage is from $\log{t}=5.00$ to 10.30 with a variable time resolution, reaching $\Delta(\log{t})=0.01$ in the youngest stellar ages. SEDs are calculated in that work for six different IMFs including the
one from \cite{fer90} which is the one used in our chemical evolution code. Therefore,
self-consistency is enforced in the sense that the same basic ingredients (IMF, stellar compositions and mean lifetimes, etc...) are used to compute the chemical evolution of the galaxy and its photometric properties.
\footnote{More details of these models can be found in \citet{mol09}}

\subsection{Star formation histories}

Let us first of all discuss the qualitative behaviour of our models and the effect of the different parameters in the context of the gas content and star formation history of the model galaxies.


\begin{figure}
\begin{center}
\includegraphics[width=8cm]{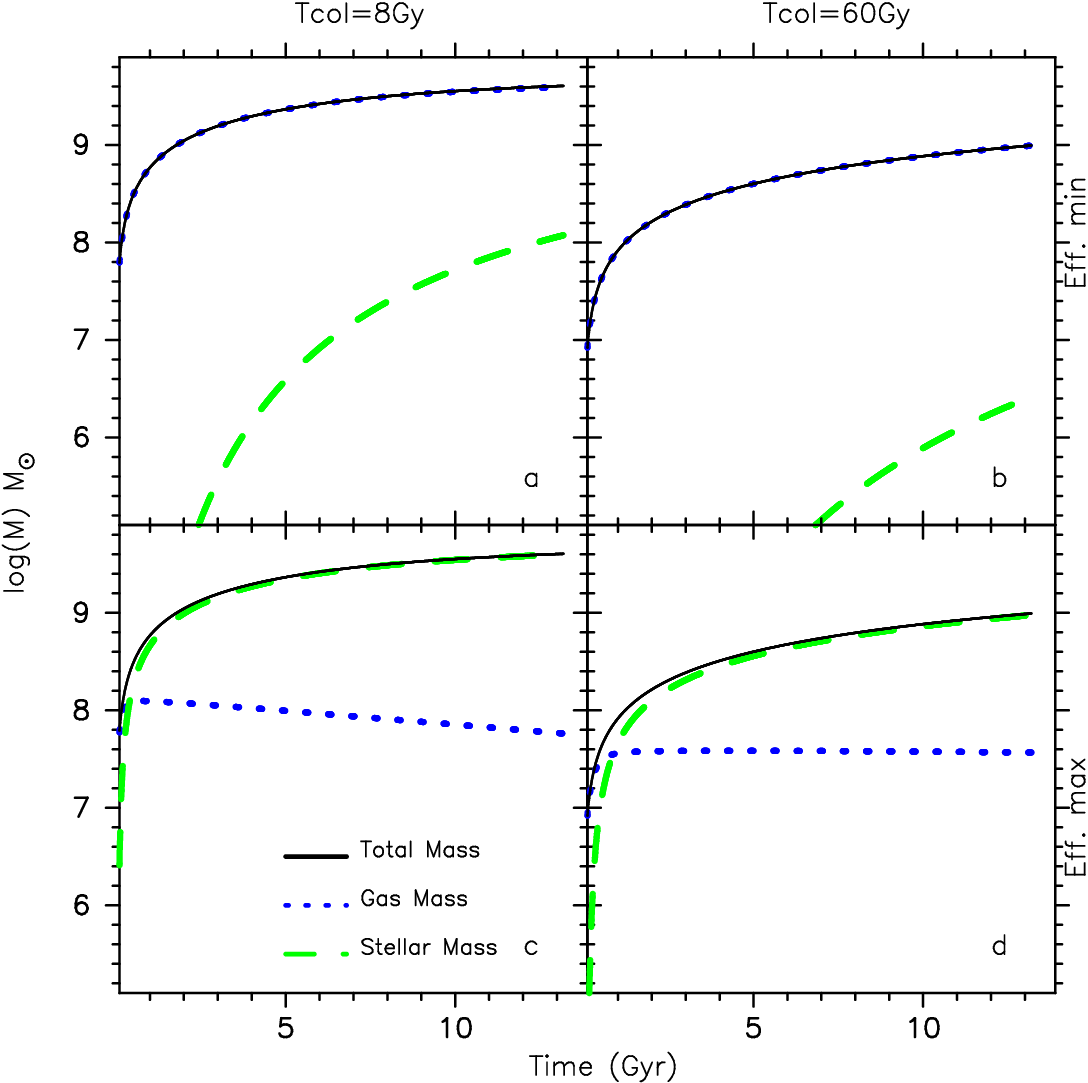}
\caption
{
Galaxy mass evolution, gas mass and stellar mass for the most massive model in its extreme cases. Panels a) and b) correspond to the lowest efficiency in star formation while panels c) and d) correspond to the highest. By columns, panel a) and c) represent the cases of minimum collapse time (maximum infall) and panels b) and d) are the opposite.
}
\label{dwMgas_t_extremos}
\end{center}
\end{figure}

In our scenario, primordial gas is continuously falling into the galaxy at a rate inversely proportional to the collapse time; for any given protogalaxy mass, longer collapse times lead to smaller total (gas+stars) masses at the present time, independent of the SFR.
On the other hand, the fraction of gas converted into stars depends both on the infall time scale and the star formation efficiency.
As an example, the evolution of the highest-mass (M6) galaxy is shown in Figure~\ref{dwMgas_t_extremos} in four extreme cases: panels a) and b) correspond to the lowest star-formation efficiency values, while the results obtained for the highest efficiencies are plotted in panels c) and d).
Looking at the figure by columns, left panels represent the cases with the shortest collapse time, while the longest ones can be seen on the right panels.
In all cases, black lines indicate the evolution of the total galaxy mass, which depends only on the collapse time.
Blue and green lines indicate the contribution of gas and stars, respectively.

In the models with minimum efficiency, the gas is the most massive component of the galaxy, and the contribution of the stellar component is so small that the total mass and the gas mass are almost the same.
In the opposite case, when efficiencies are high (bottom panels), most of the galaxy mass is in stellar form, and the black and green lines overlap. The evolution of the gas content is different in both extreme cases.
If the efficiency is low, the gas mass increases monotonically, which means that the gas infall rate is always greater than its depletion due to star formation. On the other hand, if the efficiency is high, the gas may be consumed faster than it falls, so its mass reaches a maximum at a certain moment and then becomes a decreasing function of time.

\begin{figure}
\begin{center}
\resizebox{\hsize}{!}{\includegraphics[width=8cm]{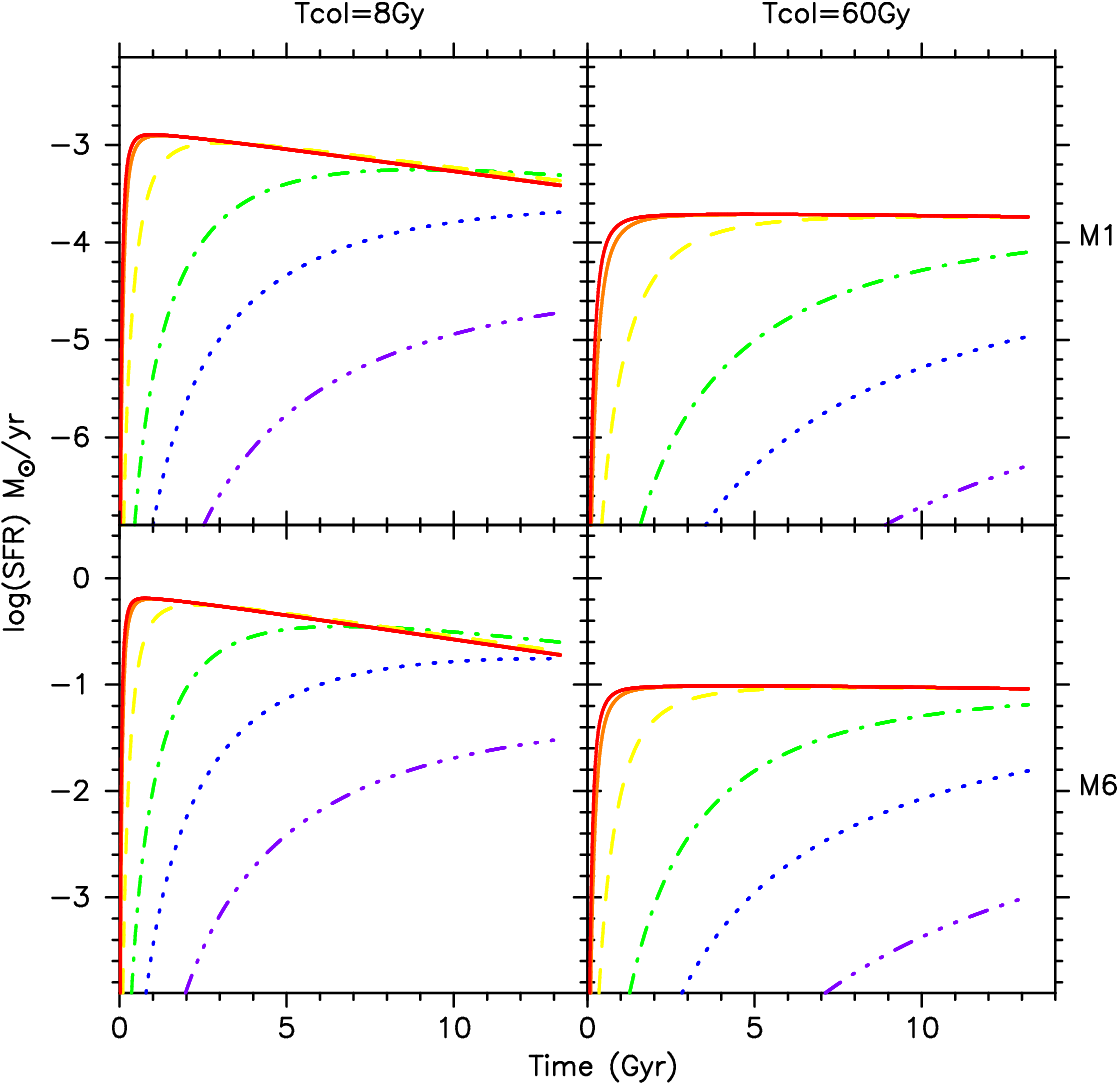}}
\caption{Star formation histories in ${\rm \msun/yr}$ for the extreme cases: lowest and highest galaxy mass and shorter and longer collapse time. Efficiency colours are: violet (dash-dot-dot-dot): very low, blue (dots): low, green (dash-dot): medium-low, yellow (dashes): medium, orange (full): high and red (full): very high. In most of the cases, the red and orange lines overlap, and both reach their maximum before the first Gyr.}
\label{SFH}
\end{center}
\end{figure}

Unlike other evolution models, where the star formation history of the galaxy is considered as an input function, in our model the SFR is related to the available gas mass through the values of the efficiency parameters.
The star formation histories for the extreme models considered in the present work are shown in Figure~\ref{SFH}.
Each panel represents one mass and one collapse time value, while different lines correspond to different efficiencies, from the highest value (in red) to the lowest (in violet).
Looking at the figure by columns, it is clear that, apart from the overall normalization, the total mass does not affect the shape of the star formation history, whose shape is almost entirely determined by the combination of the collapse time scale and the star formation efficiency.

Indeed, the main point of this plot is to show that models with high efficiency reach their maximum SFR early, and then, depending on the infall rate, the star formation rate may stay approximately constant, reaching a steady state, or decline as a consequence of the depletion of the gas reservoir.
In contrast, models with low efficiency always have enough gas to form stars.
Their SFR is always increasing, and thus it becomes maximum at the present time.
Nevertheless, it is important to remark that, even in the models with the highest efficiency, the maximum SFR is quite small compared with the values that can be reached in a typical starburst galaxy.

\section{Comparison to observational data}
\label{secResults}

We will compare the model predictions with a sample of isolated dIrr galaxies obtained from the literature \citep[mainly][]{zee00,gar02,lee03}.
\citet{zee00} selected 58 late-type, gas-rich, isolated objects with $M_{B}>-18$ from the UGC.
Gas abundances, photometry and star formation rate of these systems are available from this and subsequent works \citep{zee01,zee06,zee06a}.
\citet{gar02} compiled a sample of isolated spiral and irregular galaxies.
For the objects that were not originally observed by the author (for instance, those in common with the van Zee sample), we have consulted the original sources on a case-by-case basis.
\citet{lee03} extracted all the galaxies with well-defined distances and oxygen abundances (directly measured from the ${\rm [OIII] \lambda 4363}$ line) from the compilation of \citet{ric95}.
They also looked for new and updated data in the literature and observed five new objects with enough exposure time for the ${\rm [OIII] \lambda 4363}$ line to be measured.

Our final sample contains 61 dIrr galaxies.
Their main properties are listed in Table~\ref{tabla_dIrr}, and bibliographic references are provided in Table~\ref{ref_dIrr}.
We have ensured consistency between distances and derived HI and stellar masses.
The latter have been estimated from Popstar population synthesis models \citep{mol09} whit the same IMF than that used in our models.

In order to illustrate the general trend, we have added a representative sample of galaxies from the Sloan Digital Sky Survey \citep[SDSS DR9,][]{SDSS_9} with HI data from ALFALFA \citep{ALFALFA}.
For this set of data, N/O values have been derived by \cite{per13}, and O/H has been estimated from the [NII]/H$\alpha$ line ratio, using the calibration by \cite{per09_NO}.
Uncertainties in these calibrations are large, of the order of 0.14~dex for N/O and 0.25~dex for O/H \citep{per05}, and a shift of $-0.2$~dex has been applied to O/H in order to match the chemical abundances of the Orion nebula \citep{est09} in the N/O -- O/H plane, represented by the black star in Figure~\ref{dwNO_evol_D} below.

\begin{tiny}
\begin{table*}
\caption [Dwarf Irregular galaxies data]
{\label{tabla_dIrr}
Isolated dIrr galaxies compiled from the literature: object name, distance in Mpc, masses (neutral hydrogen, dynamical, and stellar, assuming \citet{fer90} IMF) expressed as $\log(M/\msun)$, oxygen abundance, (N/O) ratio, magnitude in Johnson $B$ band, $B-V$ colour, and SFR in \msun~yr$^{-1}$.
Superscripts denote the appropriate bibliographic reference, according to Table~\ref{ref_dIrr}. Stellar masses have been estimated from Popstar population synthesis models \citep{mol09}
}
\begin{tabular}{c|c|ccc|cc|cc|c}
\hline
Name & Distance & M(HI) & M$_{\rm dyn}$ & M$_*$ & 12 + log(O/H) & log(N/O) & M$_B$ & $B-V$ & SFR\\ \hline
CGCG007-025&$17.8^{16}$&$8.62^{18}$&$9.36^{16}$&$8.17$&$7.83\pm0.03^{16}$&$-1.48\pm0.06^{16}$&$-15.75^{16}$&$0.37\pm0.02^{16}$&$0.1150^{18}$\\
GR8-DDO155&$2.2^{16}$&$7.04^{12}$&$7.29^{16}$&$6.74$&$7.65\pm0.06^{16}$&$-1.51\pm0.07^{16}$&$-12.11^{16}$&$0.38\pm0.1^{16}$&$0.0019^{11}$\\
Haro43&$26.5^{16}$&$9.36^{2}$&$9.96^{16}$&$8.47$&$8.2\pm0.1^{16}$&$-1.41\pm0.1^{16}$&$-16.31^{16}$&$0.41\pm0.02^{16}$&$0.1116^{5}$\\
HoII&$3.39^{9}$&$8.99^{10}$&$---^{}$&$8.39$&$7.93\pm0.05^{12}$&$---^{}$&$-15.98^{12}$&$0.44^{1}$&$0.0547^{11}$\\
IC10&$0.82^{3}$&$8.14^{12}$&$9.20^{2}$&$8.45$&$8.19\pm0.15^{14}$&$-1.37\pm0.12^{14}$&$-15.85^{12}$&$0.5^{3}$&$0.0244^{11}$\\
IC1613&$0.7^{3}$&$7.97^{12}$&$8.90^{2}$&$7.97$&$7.62\pm0.05^{17}$&$-1.13\pm0.18^{17}$&$-14.33^{12}$&$0.57^{3}$&$0.0031^{11}$\\
IC2574&$4.02^{17}$&$9.16^{12}$&$---^{}$&$8.63$&$8.09\pm0.07^{17}$&$-1.52\pm0.13^{17}$&$-17.06^{12}$&$0.33^{}$&$0.0827^{11}$\\
IC4662..&$2^{17}$&$8.40^{12}$&$---^{}$&$7.85$&$8.17\pm0.04^{17}$&$-1.5\pm0.05^{17}$&$-15.84^{12}$&$0.17^{}$&$0.0657^{11}$\\
IC5152&$1.7^{3}$&$7.77^{2}$&$8.48^{2}$&$7.72$&$7.92\pm0.07^{17}$&$-1.05\pm0.12^{17}$&$-14.8^{3}$&$0.33^{3}$&$0.0169^{11}$\\
LeoA&$0.69^{16}$&$6.99^{12}$&$7.78^{2}$&$6.29$&$7.3\pm0.05^{17}$&$-1.53\pm0.09^{17}$&$-11.53^{12}$&$0.26\pm0.1^{16}$&$---^{}$\\
LMC&$0.05^{17}$&$8.82^{12}$&$9.78^{15}$&$---$&$8.37\pm0.22^{17}$&$-1.3\pm0.2^{17}$&$-17.94^{12}$&$---^{}$&$0.2441^{11}$\\
NGC1560.&$3.45^{8}$&$8.85^{12}$&$---^{}$&$8.51$&$8.1\pm0.2^{17}$&$---^{}$&$-16.37^{12}$&$0.42^{}$&$0.0434^{11}$\\
NGC1569.&$3.36^{4}$&$7.99^{12}$&$8.70^{2}$&$8.65$&$8.19\pm0.06^{12}$&$---^{}$&$-16.54^{12}$&$0.46^{}$&$0.3073^{11}$\\
NGC2366.&$3.19^{17}$&$8.95^{12}$&$9.26^{2}$&$8.40$&$7.91\pm0.05^{17}$&$---^{}$&$-16.28^{12}$&$0.38^{}$&$0.2277^{11}$\\
NGC3109&$1.25^{3}$&$8.94^{12}$&$9.82^{2}$&$8.19$&$7.73\pm0.33^{17}$&$-1.32\pm0.2^{17}$&$-15.3^{12}$&$0.48^{3}$&$0.0140^{11}$\\
NGC4214.&$2.94^{17}$&$9.24^{12}$&$9.51^{2}$&$9.07$&$8.25\pm0.1^{17}$&$-1.3\pm0.15^{17}$&$-18.04^{12}$&$0.36^{5}$&$0.1224^{11}$\\
NGC5408.&$4.81^{17}$&$8.25^{12}$&$---^{}$&$8.54$&$8\pm0.03^{17}$&$-1.46\pm0.05^{17}$&$-15.81^{12}$&$0.56^{1}$&$0.1066^{11}$\\
NGC55&$1.8^{17}$&$9.18^{12}$&$10.19^{2}$&$9.37$&$8.35\pm0.01^{17}$&$-1.44\pm0.15^{14}$&$-18.28^{12}$&$0.47^{3}$&$0.2803^{11}$\\
NGC6822&$0.49^{3}$&$8.13^{12}$&$9.21^{2}$&$8.03$&$8.11\pm0.05^{17}$&$-1.6\pm0.1^{17}$&$-14.95^{12}$&$0.47^{3}$&$0.0107^{11}$\\
Pegasus&$0.96^{3}$&$7.73^{2}$&$7.76^{2}$&$6.86$&$7.93\pm0.14^{17}$&$-1.24\pm0.15^{17}$&$-11.47^{17}$&$0.59^{3}$&$2.50E-05^{11}$\\
SextantsA&$1.47^{12}$&$8.03^{12}$&$8.60^{2}$&$7.45$&$7.54\pm0.1^{17}$&$-1.54\pm0.13^{17}$&$-14.04^{12}$&$0.35^{3}$&$0.0036^{11}$\\
SextantsB&$1.36^{12}$&$7.65^{12}$&$8.48^{2}$&$7.66$&$7.69\pm0.15^{17}$&$-1.46\pm0.06^{17}$&$-14.02^{12}$&$0.47^{3}$&$0.0013^{11}$\\
SgrdI&$1.04^{8}$&$6.22^{2}$&$6.98^{2}$&$6.59$&$7.44\pm0.2^{17}$&$-1.63\pm0.2^{17}$&$-11.6^{3}$&$0.41^{3}$&$0.0001^{11}$\\
SMC&$0.06^{12}$&$8.95^{12}$&$8.70^{2}$&$---$&$8.13\pm0.1^{17}$&$-1.58\pm0.15^{17}$&$-16.56^{12}$&$---^{}$&$0.0370^{11}$\\
UGC10445&$15.1^{16}$&$9.20^{18}$&$10.24^{16}$&$8.98$&$7.95\pm0.06^{16}$&$-1.2\pm0.1^{16}$&$-17.53^{16}$&$0.42\pm0.03^{16}$&$0.1620^{18}$\\
UGC1104&$11.1^{16}$&$8.51^{16}$&$9.31^{16}$&$8.36$&$7.94\pm0.05^{16}$&$-1.65\pm0.14^{16}$&$-16.08^{16}$&$0.4\pm0.01^{16}$&$0.0230^{18}$\\
UGC1175&$11.3^{16}$&$8.66^{16}$&$9.08^{16}$&$7.33$&$7.82\pm0.1^{16}$&$-1.5\pm0.15^{16}$&$-14.13^{16}$&$0.26\pm0.04^{16}$&$0.0032^{18}$\\
UGC11755&$18^{16}$&$8.19^{18}$&$9.76^{16}$&$8.96$&$8.04\pm0.03^{16}$&$-1.1\pm0.08^{16}$&$-17.14^{16}$&$0.5\pm0.03^{16}$&$0.0880^{18}$\\
UGC11820&$17.9^{16}$&$9.64^{2}$&$10.17^{16}$&$8.65$&$8\pm0.2^{16}$&$-1.59\pm0.1^{16}$&$-16.94^{16}$&$0.37\pm0.02^{16}$&$0.0479^{7}$\\
UGC12713&$7.5^{16}$&$8.06^{18}$&$9.29^{16}$&$7.94$&$7.8\pm0.06^{16}$&$-1.53\pm0.1^{16}$&$-14.76^{16}$&$0.46\pm0.03^{16}$&$0.0093^{18}$\\
UGC1281&$4.6^{16}$&$8.29^{18}$&$9.37^{16}$&$7.93$&$7.78\pm0.1^{16}$&$-1.29\pm0.13^{16}$&$-14.91^{16}$&$0.42\pm0.03^{16}$&$0.0084^{18}$\\
UGC12894&$7.9^{16}$&$7.96^{18}$&$8.40^{16}$&$---$&$7.56\pm0.04^{16}$&$-1.51\pm0.1^{16}$&$-13.38^{16}$&$---^{}$&$0.0033^{18}$\\
UGC191&$17.6^{16}$&$9.15^{18}$&$9.87^{16}$&$9.04$&$8.1\pm0.05^{16}$&$-1.4\pm0.1^{16}$&$-17.52^{16}$&$0.46\pm0.03^{16}$&$0.1320^{18}$\\
UGC2023&$10.2^{16}$&$8.62^{18}$&$9.23^{16}$&$8.62$&$8.02\pm0.03^{16}$&$-1.35\pm0.1^{16}$&$-16.54^{16}$&$0.44\pm0.04^{16}$&$0.0072^{18}$\\
UGC2684&$5.56^{16}$&$7.84^{16}$&$8.99^{16}$&$7.03$&$7.6\pm0.1^{16}$&$-1.38\pm0.07^{16}$&$-13.03^{16}$&$0.34\pm0.02^{16}$&$0.0009^{11}$\\
UGC290&$12.7^{16}$&$8.34^{18}$&$9.12^{16}$&$7.56$&$7.8\pm0.1^{16}$&$-1.42\pm0.11^{16}$&$-14.48^{16}$&$0.31\pm0.05^{16}$&$0.0050^{18}$\\
UGC2984&$20.6^{16}$&$9.43^{16}$&$9.99^{16}$&$9.12$&$8.3\pm0.2^{16}$&$-1.57\pm0.1^{16}$&$-18.89^{16}$&$0.2\pm0.02^{16}$&$---^{}$\\
UGC300&$19.8^{16}$&$8.61^{16}$&$---^{}$&$8.40$&$7.8\pm0.03^{16}$&$-1.5\pm0.08^{16}$&$-15.72^{16}$&$0.5\pm0.01^{16}$&$---^{}$\\
UGC3174&$7.86^{16}$&$8.43^{16}$&$9.48^{16}$&$7.85$&$7.8\pm0.1^{16}$&$-1.56\pm0.15^{16}$&$-14.76^{16}$&$0.41\pm0.02^{16}$&$0.0072^{11}$\\
UGC3647&$19.7^{16}$&$9.17^{18}$&$9.68^{16}$&$8.79$&$8.07\pm0.05^{16}$&$-1.28\pm0.1^{16}$&$-17.06^{16}$&$0.42\pm0.03^{16}$&$0.1500^{18}$\\
UGC3672&$12.7^{16}$&$8.75^{16}$&$9.25^{16}$&$7.87$&$8.01\pm0.04^{16}$&$-1.64\pm0.12^{16}$&$-15.43^{16}$&$0.27\pm0.05^{16}$&$0.0140^{18}$\\
UGC4117&$10^{16}$&$8.10^{18}$&$8.73^{16}$&$7.71$&$7.89\pm0.1^{16}$&$-1.52\pm0.15^{16}$&$-14.86^{16}$&$0.31\pm0.02^{16}$&$0.0100^{18}$\\
UGC4483&$3.21^{16}$&$7.51^{16}$&$8.11^{16}$&$---$&$7.56\pm0.03^{16}$&$-1.57\pm0.07^{16}$&$-12.55^{16}$&$---^{}$&$0.0033^{11}$\\
UGC521&$10.9^{16}$&$8.59^{16}$&$9.60^{16}$&$7.82$&$7.86\pm0.04^{16}$&$-1.66\pm0.12^{16}$&$-15.17^{16}$&$0.3\pm0.02^{16}$&$---^{}$\\
UGC5288&$5.3^{16}$&$8.23^{18}$&$9.22^{16}$&$7.81$&$7.9\pm0.03^{16}$&$-1.42\pm0.06^{16}$&$-14.44^{16}$&$0.46\pm0.06^{16}$&$0.0083^{18}$\\
UGC5716&$16^{16}$&$8.79^{16}$&$9.89^{16}$&$8.18$&$8.1\pm0.1^{16}$&$-1.59\pm0.1^{16}$&$-15.53^{16}$&$0.42\pm0.01^{16}$&$0.0234^{7}$\\
UGC5764&$7.21^{16}$&$8.14^{16}$&$8.99^{16}$&$8.28$&$7.95\pm0.04^{16}$&$-1.34\pm0.08^{16}$&$-15.3^{16}$&$0.53^{1}$&$0.0053^{11}$\\
UGC5829&$8.02^{16}$&$8.93^{16}$&$9.89^{16}$&$8.23$&$8.3\pm0.1^{16}$&$-1.76\pm0.12^{16}$&$-16.6^{16}$&$0.21^{1}$&$0.0476^{11}$\\
UGC634&$31.4^{16}$&$9.55^{16}$&$9.98^{16}$&$8.98$&$8.18\pm0.03^{16}$&$-1.58\pm0.08^{16}$&$-17.67^{16}$&$0.39\pm0.03^{16}$&$0.0730^{18}$\\
UGC6456.&$4.34^{9}$&$7.80^{12}$&$---^{}$&$---$&$7.73\pm0.05^{17}$&$-1.54\pm0.08^{17}$&$-13.9^{12}$&$---^{}$&$0.0102^{11}$\\
UGC685&$4.79^{16}$&$7.92^{18}$&$9.08^{16}$&$7.81$&$8\pm0.03^{16}$&$-1.45\pm0.08^{16}$&$-14.44^{16}$&$0.46\pm0.02^{16}$&$0.0077^{18}$\\
DDO154&$3.2^{16}$&$8.37^{16}$&$9.42^{16}$&$7.08$&$7.67\pm0.06^{16}$&$-1.68\pm0.13^{16}$&$-13.33^{16}$&$0.30.03^{6}$&$0.0032^{11}$\\
UGC8651&$3.01^{16}$&$7.63^{18}$&$8.50^{16}$&$6.97$&$7.85\pm0.04^{16}$&$-1.6\pm0.09^{16}$&$-13.2^{13}$&$0.27^{13}$&$0.0021^{11}$\\
UGC891&$10.5^{16}$&$8.67^{18}$&$9.53^{16}$&$8.25$&$8.2\pm0.1^{16}$&$-1.52\pm0.13^{16}$&$-15.53^{16}$&$0.46\pm0.03^{16}$&$0.0100^{18}$\\
DDO187&$2.5^{16}$&$7.32^{12}$&$8.00^{2}$&$6.71$&$7.75\pm0.05^{16}$&$-1.8\pm0.12^{16}$&$-12.63^{16}$&$0.25\pm0.05^{16}$&$0.0001^{18}$\\
UGC9240&$2.79^{16}$&$7.90^{18}$&$9.17^{16}$&$7.46$&$7.95\pm0.03^{16}$&$-1.6\pm0.06^{16}$&$-13.96^{16}$&$0.37\pm0.03^{16}$&$0.0035^{18}$\\
UGC9992&$8.6^{16}$&$8.28^{18}$&$8.44^{16}$&$7.86$&$7.88\pm0.12^{16}$&$-1.26\pm0.19^{16}$&$-14.97^{16}$&$0.37\pm0.04^{16}$&$0.0062^{18}$\\
UGCA20&$8.65^{16}$&$8.25^{16}$&$9.34^{16}$&$7.40$&$7.6\pm0.1^{16}$&$-1.56\pm0.15^{16}$&$-14.13^{16}$&$0.3\pm0.01^{16}$&$0.0107^{11}$\\
UGCA292&$3.1^{16}$&$7.60^{18}$&$7.93^{16}$&$5.92$&$7.32\pm0.06^{16}$&$-1.44\pm0.1^{16}$&$-11.43^{16}$&$0.08\pm0.1^{16}$&$0.0016^{11}$\\
UGCA357&$16.5^{16}$&$8.91^{16}$&$9.91^{16}$&$8.15$&$8.05\pm0.05^{16}$&$-1.55\pm0.11^{16}$&$-15.5^{16}$&$0.41\pm0.01^{16}$&$---^{}$\\
WLM&$0.92^{17}$&$7.79^{12}$&$8.60^{2}$&$7.86$&$7.77\pm0.1^{17}$&$-1.46\pm0.05^{17}$&$-13.92^{12}$&$0.6^{3}$&$0.0017^{11}$\\
\hline
\end{tabular}
\end{table*}
\end{tiny}

\subsection{Chemical abundances}


Due to galactic evolution, there should exist a relation between the metallicity and the amount of mass in stars and gas. If the system is dominated by gas, one can expect that the metallicity should be very low, because there have not been enough stars to enrich the interstellar medium. In contrast, if the stellar fraction is high, the metallicity should also be high. In a scenario with instantaneous recycling, metallicity-independent yields, and no mass exchange, usually called the simple \emph{closed-box} model (SCBM), the relation between metallicity and gas fraction is approximated when $Z \ll 1$ by
\begin{equation}
 Z = y\ \ln \mu^{-1}
 \label{eq_Z_mu_CBM}
\end{equation}
where $Z$ is the metallicity, usually expressed in terms of the abundance of a primary element such as oxygen, $y$ is the corresponding net stellar yield, and
\begin{equation}
\mu=\frac{M_{\rm gas}}{M_{\rm gas}+M_{\rm stars}}
\label{eq_mu_bar}
\end{equation}
is the gas fraction, defined as the ratio between the gas and the total \emph{baryonic} mass
\begin{equation}
 M_{\rm gas} \approx 1.34\ M_{\rm HI}
\end{equation}
where $M_{\rm HI}$ is the neutral hydrogen mass, measured from the flux of the 21~cm line, with the factor $1.34$ accounting for the presence of helium, and the mass in other (e.g. molecular or ionized) hydrogen phases is not included because CO is not observationally detected in most of these objects.

Historically, the discrepancy between the observed relation between metallicity and gas fraction with respect to the predictions of the SCBM has been one of the most important arguments for the need for galactic winds in dwarf galaxies \citep[see e.g.][and references therein]{pei82}.
More specifically, different wind efficiencies are advocated as the most natural explanation for the \emph{scatter} in the $Z-\mu$ relation \citep{mat83} (see however Appendix~\ref{secMuDyn}).

\begin{figure}
\begin{center}
\includegraphics[width=8cm]{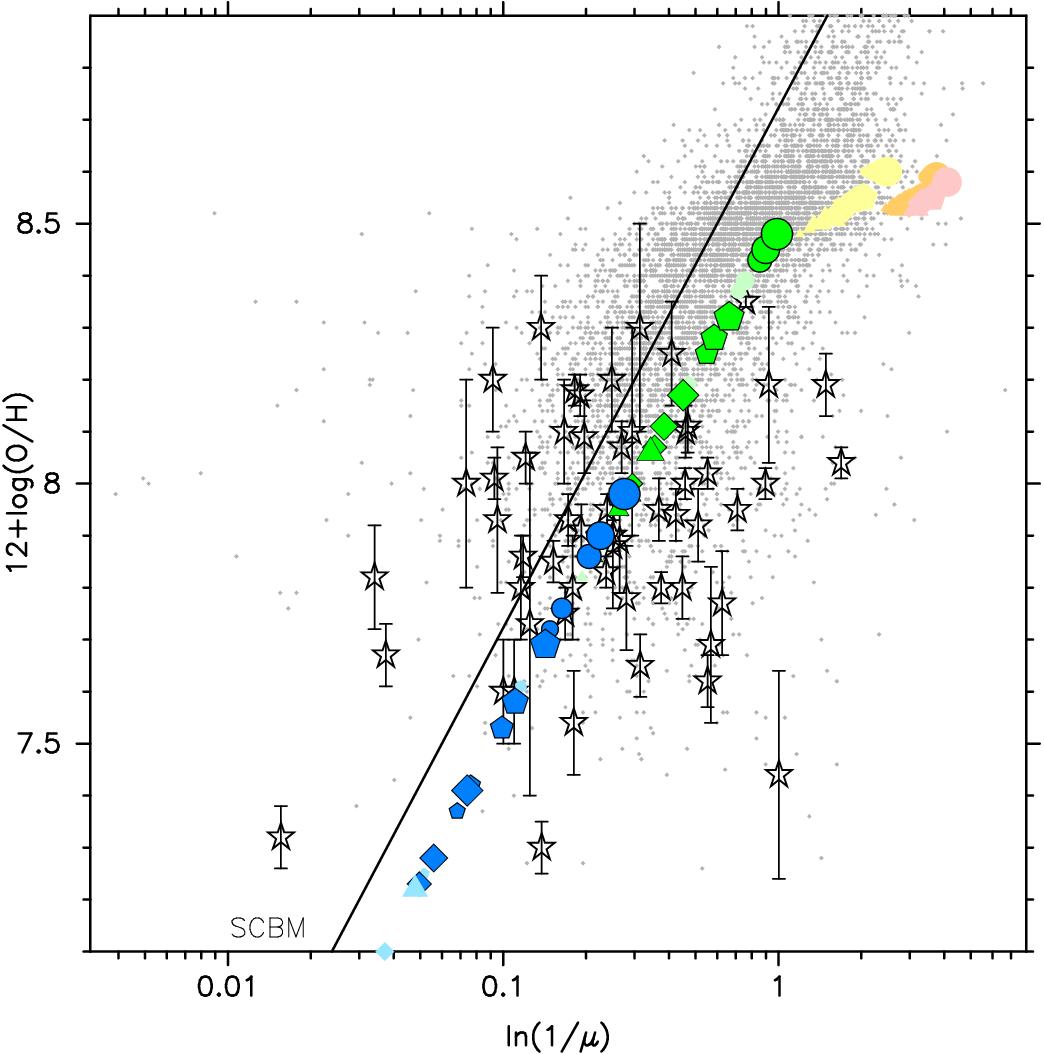}
\caption
{
Gas fraction-metallicity relation.
Symbol size represents galaxy mass, colours (violet,  dark blue, light green, yellow, orange, and red) indicate increasing star formation efficiency (Table~\ref{tabla_Efficiencies}), and shapes (circles, pentagons, squares, and triangles) refer to infall times (8, 20, 40 and 60~Gyr, respectively).
Colour shading has been used to highlight models compatible with dIrr data (see discussion in \ref{sec_fotometria}).
A SCBM with $y_{\rm eff}=0.006$ is plotted as a solid line, and stars with error bars correspond to the observations listed in Table~\ref{tabla_dIrr}.
Small grey dots correspond to SDSS+ALFALFA data. 
}
\label{dwOH_mu}
\end{center}
\end{figure}

Figure~\ref{dwOH_mu} shows the gas metallicity, in terms of 12+log(O/H), versus the baryonic gas fraction $\mu$ for the objects in Table \ref{tabla_dIrr}. The results of our model grid are represented by points with different sizes, colours, and shapes, that encode the different values of the model parameters. In general, models with low star formation efficiencies provide a good match to the observational data, especially if we take into account that the errors in the stellar and gas masses may be quite large.

\begin{table}
\begin{center}
\caption{\label{ref_dIrr}References for the observational data in Table~\ref{tabla_dIrr}.}
\begin{tabular}{|c|l|}

\hline
 &References\\
\hline
1&\cite{RC3}\\
2&\cite{fer00}\\
3&\cite{gar02}\\
4&\cite{gro08}\\
5&\cite{hun06}\\
6&\cite{hun10}\\
7&\cite{Jam04}\\
8&\cite{kar03b}\\
9&\cite{kar04}\\
10&\cite{kar07}\\
11&\cite{ken08}\\
12&\cite{lee03}\\
13&\cite{mak09}\\
14&\cite{mat98}\\
15&\cite{mat83}\\
16&\cite{zee06a}\\
17&\cite{zee06}\\
18&\cite{zee01}\\
\hline
\end{tabular}
\end{center}
\end{table}

One can readily see that observational data may be reproduced by using the naive prediction of the SCBM, equation~(\ref{eq_Z_mu_CBM}), as a fitting function in terms of an \emph{effective} oxygen yield.
The black solid line corresponds to $y_{\rm eff}=0.006$, close to the true oxygen yield \citep[within a factor of 2 uncertainty, see][page 239]{pag09}.
Our models follow a similar trend, somewhat shifted to the right, corresponding to a slightly lower effective yield, and they follow a tight sequence on the $Z-\mu$ plane, regardless of the precise values of the infall timescale or the star formation efficiency.

According to Figure~\ref{dwOH_mu}, current data seem to indicate that our infall scenario may accommodate the observed $Z-\mu$ relation.
This, of course, does not mean that the infall of primordial gas is the only possible explanation.
It is important, though, that the baryonic gas fraction $\mu$ -- rather than $\mu_{\rm dyn}$ -- is considered, and that the theoretical models (including infall, outflow, or both) are able to reproduce not only the average $Z-\mu$ relation, but also its low intrinsic scatter (contrary to the large dispersion obtained when $\mu_{\rm dyn}$ is used; see Appendix~\ref{secMuDyn}).


\begin{figure}
\begin{center}
\includegraphics[width=8cm]{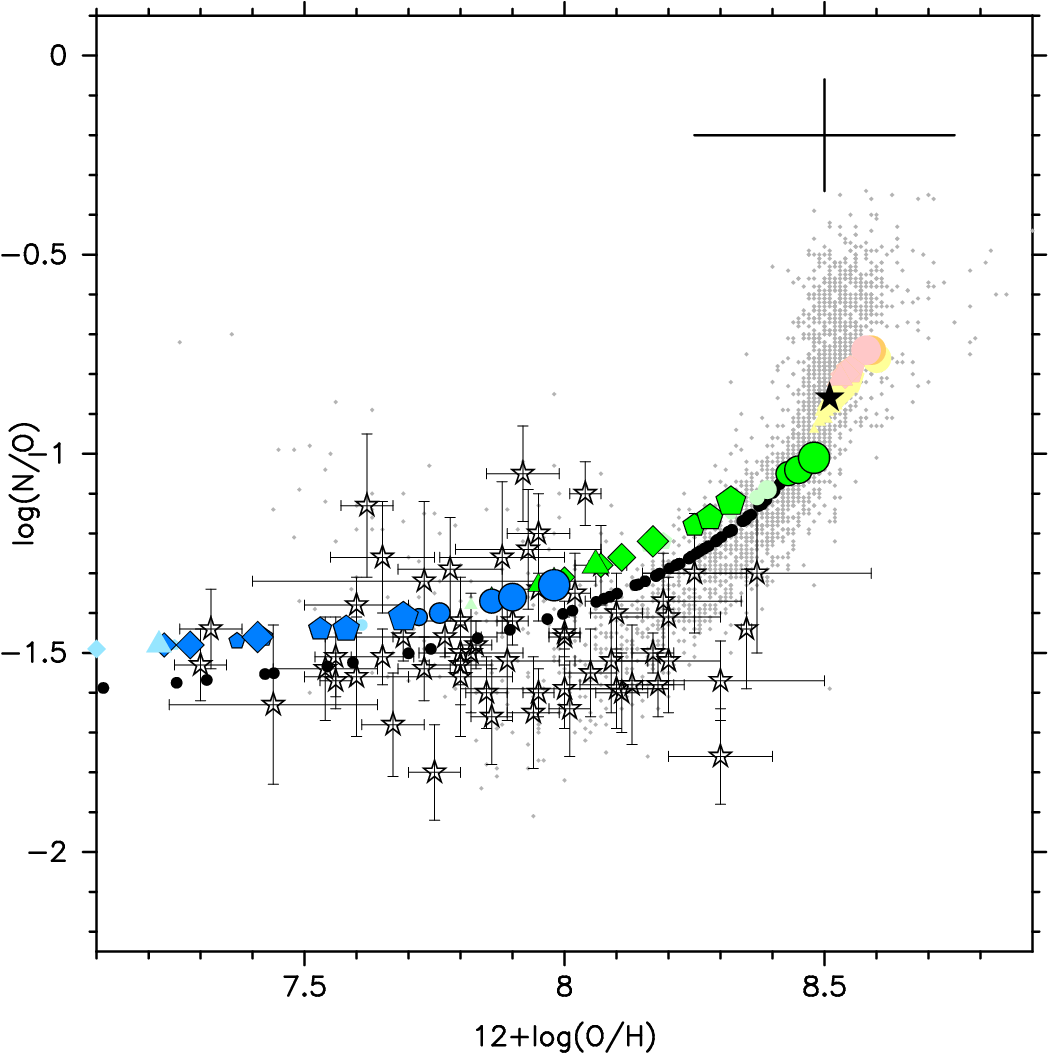}
\caption{N/O ratio as a function of gas oxygen abundance.
Colour symbols mark the final values attained by our evolutionary models (shapes, sizes, and colours have the same meaning as in Figure~\ref{dwOH_mu}). 
Black circles show the (N/O)--(O/H) relation for models at 1 Gyr.
Open stars represent the dIrr galaxies in Table~\ref{tabla_dIrr}, and small dots correspond to SDSS+ALFALFA data. Estimated error bars are shown in the upper right corner of the figure. Solid star illustrates the chemical abundances measured in the Orion nebula \citep{est09}.
}
\label{dwNO_evol_D}
\end{center}
\end{figure}

The relation between N/O and oxygen abundance is illustrated in Figure~\ref{dwNO_evol_D}, where colour symbols correspond to the present-time abundances for our models.
As pointed out by \citet{mol06}, any galaxy that is allowed to evolve for more than one Gyr (shown by small black circles) will end up in a very narrow locus on the N/O -- O/H plane.
The exact location along this sequence will be set by the model parameters (mainly the collapse time and the star formation efficiency).
In this scenario, all galaxies should follow a well-defined N/O relation, which is a robust \textit{prediction} of our models, unless their chemical abundances are dominated by a very young stellar population.

In general, the observational data from our compilation of dIrr galaxies (Table~\ref{tabla_dIrr}), represented by stars with error bars, are braoadly consistent with model predictions, albeit with a large scatter.
These data constitute the low mass and metallicity end of a sequence that becomes more evident when the objects from the SDSS+ALFALFA sample are added.
The sequence defined by the data seems to lie below and/or to the right of the predicted one.
Possible causes for this offset could be found in combinations of IMF and stellar nucleosynthesis different from those assumed in our models.
We have explored correlations with the star formation activity (SFR, SSFR, gas consumption timescale), but we have not found any obvious systematic trend.

On the other hand, galaxies with high N/O and low oxygen abundance may be explained either by nitrogen enrichment (due to e.g. Wolf-Rayet stars) or by the effect of selective galactic winds.
In the first case, Wolf-Rayet stars lose their envelope during the early phases of their evolution (ages $< 3$~Myr), altering the pattern of CNO abundances in the stellar cluster where they reside, and the nitrogen yield may increase up to an order of magnitude with respect to the value produced by normal supernova explosions \citep{mol12}.
In the second case, oxygen is depleted by the galactic wind.
When low-mass stars die, they find an oxygen-poor ISM, and thus the ratio N/O would be large.
Under this scenario, the (few) objects located in the upper-left area of the plot, well above the predicted N/O relation, would be good candidates to be studied with evolutionary models including enriched winds \citep{per03}.

In any case, there is substantial scatter, specially at low metallicities.
The spread of the observed points in the N/O -- O/H plane has been previously discussed on the basis of different SFR in different galaxies \citep{pil92,pil93}, stochastic IMF sampling \citep{car08c}, episodic gas infall \citep{koe05} and inhomogeneus chemical evolution \citep{kar05,ces08}. 
We would like to point out, however, that the observed scatter is not totally inconsistent with the error bars.

\subsection{Photometry}
\label{sec_fotometria}

The relations between luminosity, total mass, and metallicity provide important clues about the evolution of dwarf galaxies. 
For instance, it has long been known \citep{leq79} that the highest-mass galaxies have the highest metallicities, suggesting that they are old and/or have been efficient in forming stars.

\begin{figure}
\begin{center}
\includegraphics[width=8cm]{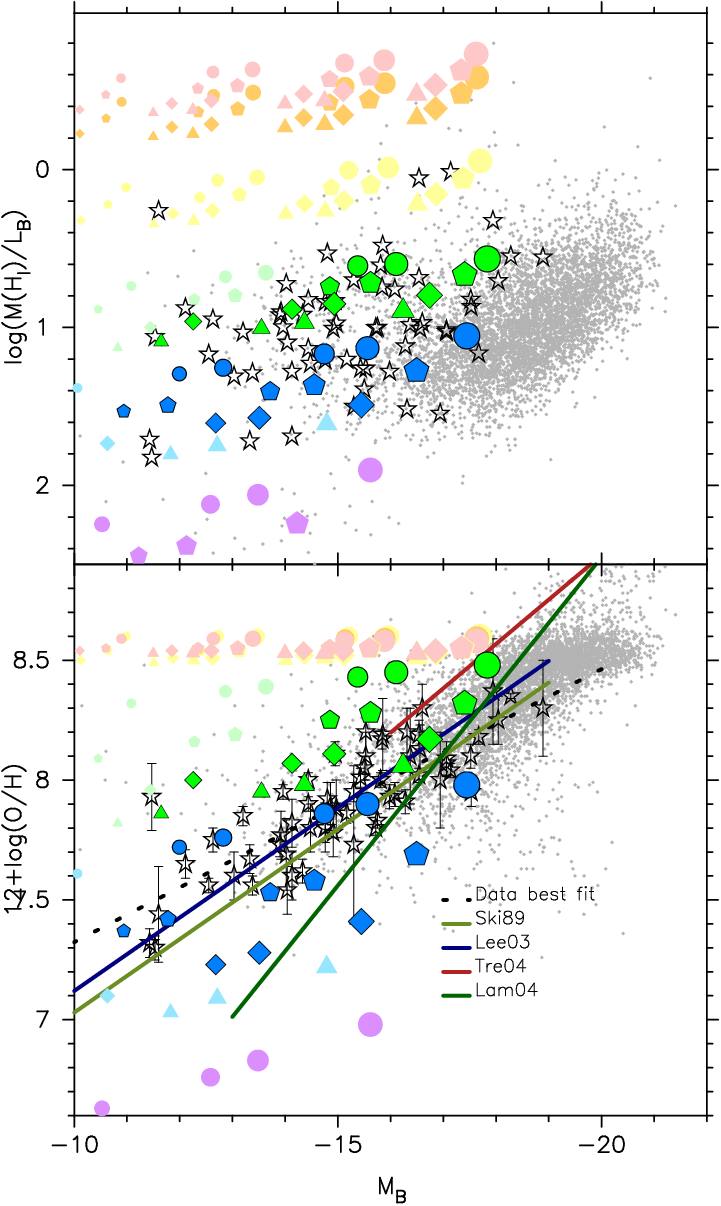}
\caption{
Top panel: gas mass-luminosity relation. Colours and symbols have the same meaning as in previous figures.  In particular, colour shading has been used to highlight models compatible with dIrr data (see text).
Note that the vertical axis is inverded for easier comparison with bottom panel.
Bottom panel:  Relation between luminosity and gas oxygen abundance. 
Lines are form \citet[Ski89]{ski89} and \citet[Lee03]{lee03}, derived for dIrr galaxies, as well as \citet[Tre04]{tre04} and \citet[Lam04]{lam04}, for field galaxies.
The dotted line corresponds to the best fit for the data in Table~\ref{tabla_dIrr}.
}
\label{dwMHILb}
\end{center}
\end{figure}

Figure \ref{dwMHILb} shows the gas mass and oxygen abundance as a function of stellar luminosity.
For an easier comparison, the vertical axis of the top panel has been inverted, so that the upper area of both panels corresponds to high-efficiency models.

In the top panel of the figure, the effects of mass and efficiency appear neatly separated.
For a given efficiency, model galaxies with a higher mass in stars are brighter in the B band and show an almost constant gas mass-to-light ratio.
Objects that have been more efficient in forming stars, and hence have much less gas remaining for a given amount of stars, occupy the top of the plot.
In other words, the B magnitude roughly measures the stellar mass, while the gas mass-to-luminosity ratio measures the star formation efficiency.
It can be seen that, while the observational data cover a broad range in luminosity (M$_B$ between -12 and -20), they cover only a restricted range in star formation efficiency, from low to moderate values.

The bottom panel shows the relation between luminosity and gas oxygen abundance.
In agreement with previous studies, our compilation of observational data displays a good correlation between metallicity and luminosity.
Some of the observational relations quoted in the literature have been over-plotted.
Models with recent star formation (low and very low efficiency) are located below the observed Z/L relation, while those above it correspond to galaxies with higher star formation in the past, older stellar populations, and high oxygen abundances (see Figure~\ref{SFH}).
The dispersion at a given B luminosity can be understood as due to the different star formation histories of the galaxies.

\begin{figure}
\begin{center}
\includegraphics[width=8cm]{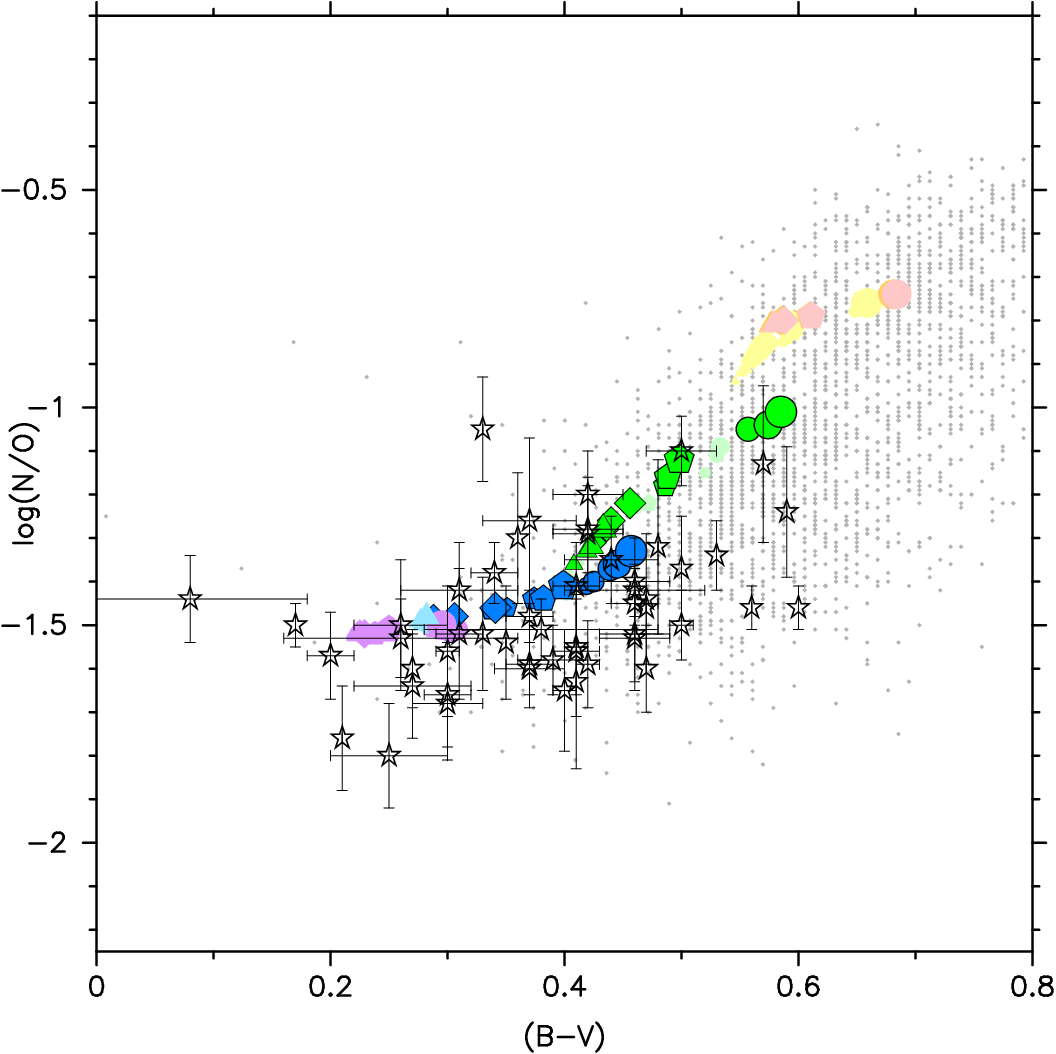}
\caption{Relation between colour and nitrogen-to-oxygen ratio. Colours and symbols have the same meaning as in previous figures.}
\label{dwBV_NO}
\end{center}
\end{figure}

In addition, the N/O ratio is represented in Fig.~\ref{dwBV_NO} as a function of B-V colour. Both models and data show a well-defined trend, in the sense that higher N/O values correspond to redder galaxies, as already pointed out by \citet{zee06}. 

Physically, the relation between metallicity and mass (or luminosity) can be explained in three possible ways: 1) Low-mass, low-metallicity galaxies are younger (i.e. they started forming stars more recently) than more massive and metal-rich systems. 2) Dwarf galaxies are as old as the bigger ones, but they form less stars for the same amount of gas at all times (i.e. the lower the galaxy mass, the lower the star formation efficiency). 3) If the same amount of stars are formed per unit gas mass, independent of total galaxy mass, enriched galactic winds may decrease the metallicity in dwarf galaxies while maintaining a gas fraction compatible with observations.

In our scenario, all galaxies form stars throughout their whole lives, and there are no outflows from the system.
Therefore, both the mass/luminosity-metallicty relation and the sequence in the N/O-(B--V) plane must arise from differences in the collapse time and/or the star formation efficiency.
The observational data in Figures~\ref{dwMHILb} and~\ref{dwBV_NO} can be reproduced by our models assuming appropriate values of these parameters, which are highly degenerate.

As in previous figures, we have shaded in a lighter colour those models whose predictions are outside the range encompased by our present dataset (i.e. those found by visual inspection to be unable to reproduce the $M_{B} - M(HI)$ or mass-metallicity relations).
As it can be readily seen, successful models feature low to moderate star formation efficiencies.
Moreover, there is some evidence for `downsizing', suggesting that more luminous galaxies formed  a larger fraction of their stars in the past (i.e. shorter collapse timescale and/or higher star formation efficiency), in agreement with previous analyses of the M-Z relation \citep[e.g.][]{bro07,cal09}.

\begin{figure}
\begin{center}
\includegraphics[width=8cm]{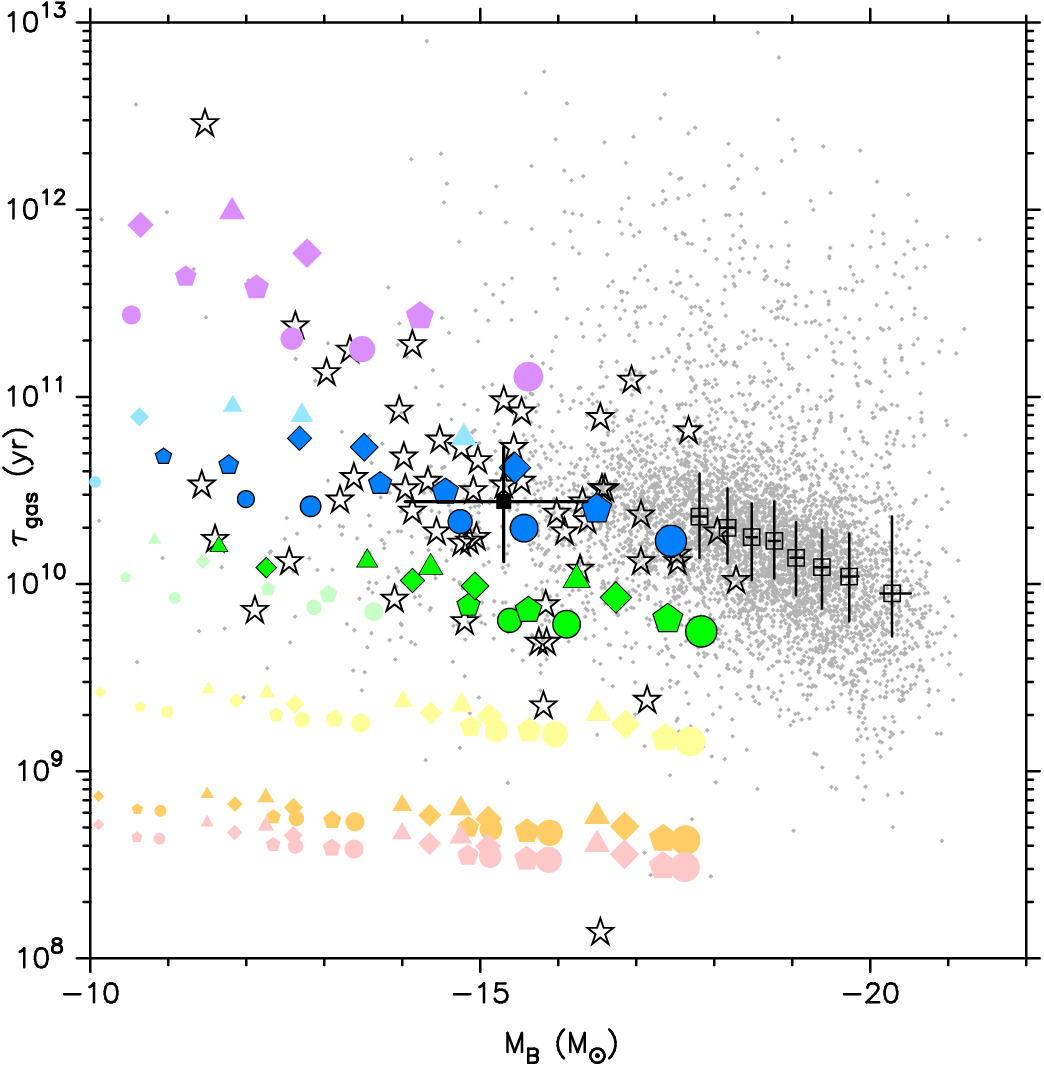}
\caption{Star formation efficiency (in terms of gas consumption timescale) as a function of B magnitude. Open squares represent the median values of SDSS+ALFALFA data, grouped in bins of  500 galaxies, while the filled square corresponds to the median values for galaxies in Table~\ref{tabla_dIrr}. Other colours and symbols have the same meaning as in previous figures.}
\label{dwTau_gas}
\end{center}
\end{figure}

We address the question of whether higher-mass galaxies form their stars more efficiently in Figure~\ref{dwTau_gas}, where the gas consumption timescale
\begin{equation}
\tau_{\rm gas} \equiv \frac{ M_{\rm gas} }{ SFR }
\end{equation}
is represented as a function of $M_{B}$. Observational data do indeed show a weak trend of increasing efficiency (lower $\tau_{\rm gas}$) towards brighter objects.
To make this trend more evident, we have sorted the SDSS+ALFALFA data in order of increasing luminosity and divided the sample in uniform bins containing 500 galaxies each, starting at $M_B = -17$ to avoid selection effects. For each bin, we have computed the median mass and gas consumption timescale and show them as open squares (error bars mark the first and third quartiles).
The median values for all the dIrr in Table~\ref{tabla_dIrr} have been plotted as a solid square. Once again, the observations are best described by models with low and moderate efficiencies.

\section{Conclusions}
\label{secConclusions}

We have investigated a self-consistent kind of models of the formation and evolution of gas-rich dwarf galaxies that considers infall of primordial, external gas, but does not include any outflow or galactic wind.
A grid of chemical evolution and spectrophotometric models has been developed with different efficiencies, infall timescales, and total masses.
The star formation rate is continuous and smooth, and it is related to the gas content by means of an efficiency parameter.
In this scenario, gas accretion and cooling up to the atomic phase are assumed to take place in a characteristic time $\tau$, whereas molecular clouds and stars form according to an overall efficiency $\epsilon$.
The precise values of $\tau$ and $\epsilon$ are set by very complex, yet to be determined, physics, and we treat them as free parameters that may vary from one galaxy to another.

These models are intended to represent field dwarf galaxies with recent star formation but not bursting (i.e. dIrr galaxies), and their predictions have been compared with a selection of suitable observations compiled from the literature.
As a result of this comparison, it has been found that:

\begin{enumerate}
\item
The infall scenario -- formation by continuous accretion of primordial gas -- is able to reproduce the gas fractions, chemical abundances, and photometric properties observed in most galaxies in our sample, without resorting to either galactic winds or selective mass loss.
We thus conclude that gas-rich dwarf galaxies with moderate to low star formation rates, without bursts, may be able to retain the vast majority of their gas and metals.
\item
More specifically, the predicted ratio between gas and stars, as well as its relation with galaxy metallicity, are in fair agreement with the values observed in most systems.
\item
The relative abundances of nitrogen and oxygen are particularly interesting.
Our models predict tight relations in the N/O -- O/H and N/O -- (B-V) planes, in rough agreement with observations (albeit the latter showing a large spread).
The objects that show higher N/O than predicted by the models may be good candidates for selective gas loss due to enriched galactic winds.
Alternatively, they may host Wolf-Rayet stars, which lose mass by stellar winds and pollute the ISM with nitrogen.
\item
Combining the relations between gas mass, nitrogen and oxygen abundances, B-band luminosity and B-V colour, we conclude that smaller galaxies must form fewer stars per unit gas mass than larger systems.
Models with very low or high efficiency do not match any object in our sample.
\end{enumerate}

The fact that low-mass galaxies are less efficient in forming stars may be interpreted in terms of a more important impact of supernova feedback.
However, other alternative explanations are possible: here, we suggest a scenario where SN explosions and/or other mechanisms lowers the efficiency of star formation in low-mass objects, but they do not drive the gas out of the galaxy.
Our results show that this scenario is indeed compatible with the available data, reinforcing the conclusions reached by \citet{tas08} that ``the star formation law governing the conversion of cold gas into stars, rather than SN-driven outflows, is the dominant factor in shaping properties of faint galaxies''.
Nevertheless, it is important to remark that, even in the models with the highest efficiency, the maximum SFR is quite small compared with the values that can be reached in a burst.
In this sense, all our galaxies could be considered as relatively quiescent.

Our conclusions do not depend on the exact physical process (e.g. injection of thermal energy, linear momentum, angular momentum, non-thermal pressure, photoionization) nor the agent (stars, AGN, magnetic fields...) that is advocated to regulate star formation in any particular model.
This is especially relevant in the context of numerical simulations, where supernova feedback is quite often the main mechanism that sets the star formation efficiency in dwarf galaxies.
If the contribution of other processes were not negligible (which arguably seems quite likely), the amount of supernova feedback would be overestimated.
This is, in fact, a good example of why a very simple, very general model is useful: if the main mechanism that is responsible for regulating the accretion of gas (or the formation of molecular clouds, or the formation of stars) is, say, the magnetic pressure or the microscopic gas turbulence, the required amount of supernova feedback would be much smaller (and winds would be much weaker) than predicted by many `first-priciples' models.

Nevertheless, we would like to finish this article with a word of caution.
According to our models, dwarf galaxies that convert their gas into stars efficiently, if they exist, would have exhausted most of their gas quickly.
Therefore, they would display faint luminosities and red colours, and they would have never been included into the data body.
Whether such objects may bear a relation with gas-poor dwarf galaxy types, like dE or dSph, will be investigated in a future work.


 \section*{Acknowledgments}

We thank the anonymous referee for a thorough, insightful, and constructive report, that has contributed to greatly improve the clarity of the paper.

This work has been financially supported by the Spanish Government projects ESTALLIDOS: AYA 2007-67965-C03-02, AYA 2007-67965-C03-03, AYA2010-21887-C04-03, AYA2010-21887-C04-04 and Consolider-Ingenio 2010 Program grant CSD2006-00070: First Science with the GTC\footnote{http://www.iac.es/consolider-ingenio-gtc}.
Partial support by the Comunidad de Madrid under grants S-0505/ESP/000237 (ASTROCAM) and CAM S2009/ESP-1496 (AstroMadrid) is also acknowledged.
YA is funded by the \emph{Ram\'{o}n y Cajal} programme (contract RyC-2011-09461), now managed by the \emph{Ministerio de Econom\'{i}a y Competitividad} (fiercely cutting back on the Spanish scientific infrastructure).

Funding for SDSS-III has been provided by the Alfred P. Sloan Foundation, the Participating Institutions, the National Science Foundation, and the U.S. Department of Energy Office of Science. The SDSS-III web site is http://www.sdss3.org/.

SDSS-III is managed by the Astrophysical Research Consortium for the Participating Institutions of the SDSS-III Collaboration including the University of Arizona, the Brazilian Participation Group, Brookhaven National Laboratory, University of Cambridge, Carnegie Mellon University, University of Florida, the French Participation Group, the German Participation Group, Harvard University, the Instituto de Astrofisica de Canarias, the Michigan State/Notre Dame/JINA Participation Group, Johns Hopkins University, Lawrence Berkeley National Laboratory, Max Planck Institute for Astrophysics, Max Planck Institute for Extraterrestrial Physics, New Mexico State University, New York University, Ohio State University, Pennsylvania State University, University of Portsmouth, Princeton University, the Spanish Participation Group, University of Tokyo, University of Utah, Vanderbilt University, University of Virginia, University of Washington, and Yale University.


\bibliographystyle{mn2e} 


\appendix

 \section{Pitfalls of the dynamical mass and the $Z-\mu$ relation}
 \label{secMuDyn}

\begin{figure}
\begin{center}
\includegraphics[width=8cm]{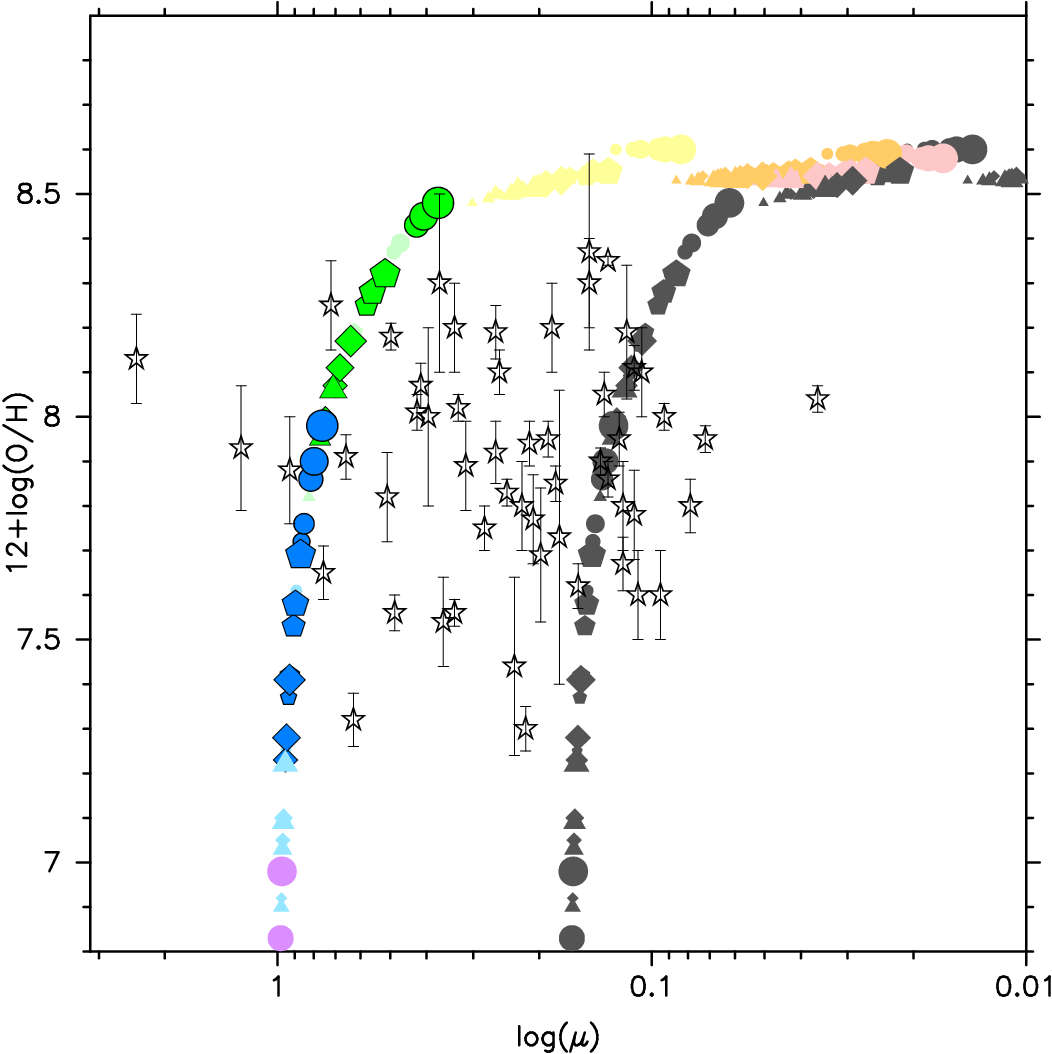}
\caption
{
Dynamical gas fraction~(eq. \ref{eq_mu_dyn}) \textit{vs} metallicity relation.
Models without dark matter (same as those on the fig~\ref{dwOH_mu}) are shown in light grey; a dark matter contribution equal to the cosmic value $\Omega_{\rm dm}/\Omega_{\rm b} \simeq 5$ (arguably an upper limit) is indicated by the dark symbols.
}
\label{dwOH_mu_mal}
\end{center}
\end{figure}

As noted by e.g. \citet{gar08}, some authors use the total dynamical mass
\begin{equation}
\mu_{\rm dyn} = \frac{M_{\rm gas}}{M_{\rm dyn}} = \frac{M_{\rm gas}}{M_{\rm gas}+M_{\rm stars}+M_{\rm dm}}
\label{eq_mu_dyn}
\end{equation}
instead of the baryonic mass when discussing the gas fraction-metallicity relation in the context of galactic outflows.
It is important to note that the dynamical mass includes the dark matter component within the observed region, which does \emph{not} participate in chemical evolution.
We have represented in Figure~\ref{dwOH_mu_mal} the metallicity vs gas fraction relation that would be derived if the dynamical mass, including dark matter, was \textit{incorrectly} considered in the definition of the gas fraction.
The contribution of the dark matter halo to the total mass budget is rather uncertain, especially in the inner, luminous part of galaxies, and  it may constitute an important source of scatter.
In principle, a reasonable upper limit would be the cosmic ratio \citep[of the order of 5 for a $\Lambda$CDM universe; see e.g.][]{kom11}, whereas the lower limit would simply be a baryon-dominated object, where the amount of dark matter is negligible.
The actual value of $\mu_{\rm dyn}$ will be determined by the detailed properties of the galaxy under study, as well as the radius within which $M_{\rm dyn}$ is measured, but it will most likely be bracketed by
\begin{equation}
\mu/6 \le \mu_{\rm dyn} \le \mu
\end{equation}
For the sake of comparison, the model results are now plotted for the extreme cases $M_{\rm dm} = 0$ and $M_{\rm dm} = 5\, ( M_{\rm gas} + M_{\rm stars} )$.

\end{document}